\documentclass[twocolumn,superscriptaddress,secnumarabic,amssymb, nobibnotes, aps, prl]{revtex4-1}
\usepackage{graphicx}
\usepackage{amsfonts}
\usepackage{amssymb}
\usepackage{amsmath}
\usepackage{psfrag}
\usepackage{natbib}
\usepackage{mathtools}
\usepackage[dvipsnames]{xcolor}

\usepackage{ulem}
\usepackage{float}
\usepackage[colorlinks=true, linkcolor=blue, urlcolor=blue, citecolor=black]{hyperref}
\usepackage{caption}
\newcommand{\beqa}{\begin{eqnarray}}
\newcommand{\eeqa}{\end{eqnarray}}
\newcommand{\beq}{\begin{equation}}
\newcommand{\eeq}{\end{equation}}

\usepackage[justification=justified,singlelinecheck=false]{caption}

\begin{document}


\title{Contact-mediated asymmetric delay and rate memory in snapping beam pairs}

\author{Gavin Linton}
 \altaffiliation{These authors contributed equally to this work.}
 \affiliation{Department of Mechanical \& Aerospace Engineering, Syracuse University, Syracuse, NY 13244, USA}

\author{Shima Mahmoudirad}
 \altaffiliation{These authors contributed equally to this work.}
 \affiliation{Department of Mechanical \& Aerospace Engineering, Syracuse University, Syracuse, NY 13244, USA}
 
\author{Nidhi Pashine}
 \affiliation{Department of Physics, Syracuse University, Syracuse, NY 13244, USA}
  \affiliation{BioInspired Syracuse: Institute for Material and Living Systems, Syracuse University, Syracuse, NY 13244, USA}
\author{Anupam Pandey}
\email{apande05@syr.edu}
 \affiliation{Department of Mechanical \& Aerospace Engineering, Syracuse University, Syracuse, NY 13244, USA}
  \affiliation{Department of Physics, Syracuse University, Syracuse, NY 13244, USA}
 \affiliation{BioInspired Syracuse: Institute for Material and Living Systems, Syracuse University, Syracuse, NY 13244, USA}

\begin{abstract}
\noindent Slender structures in dense environments interact through contact, which alters how and when they become unstable. Here we show that contact makes snap-through history dependent, so that when and whether a beam pair snaps depends on how long the contact is loaded. Under cyclic compression, the pair snaps on loading, snaps on unloading, or stays packed, with the loading and unloading rates selecting among the three. Upon contact, one beam rolls along the other without slip, and faster driving lets it roll further before it slides and snaps: to larger compression on loading and to smaller compression on unloading. This delay is asymmetric between loading and unloading as the unloading branch carries a memory of the loading rate. Interestingly, above a threshold compression the pair snaps even when the compression is stopped, with the time to snap diverging at that threshold. By exploiting this asymmetric, rate-dependent delay, we program ordered sequences of snap events in arrays of pairs with tunable gaps, and demonstrate that the energy dissipated per cycle becomes a two-parameter function of the loading and unloading rates, increasing fourfold when, at the fastest loading, the return is slowed.

  
\end{abstract}

\maketitle

\section*{Introduction}
Snap-buckling is a prototypical mechanical instability in slender elastic structures, characterized by a sudden transition between stable configurations once a control parameter crosses a critical threshold~\cite{vella2019buffering, pandey2014dynamics, gomez2017critical, sano2018snap}. This instability enables rapid actuation, mechanical amplification, and energy release, as well as drives switches, metamaterials, and bistable devices~\cite{wang2023insect, gomez2017passive, drotman2021electronics,rothemund2018soft, rafsanjani2015snapping}. In isolated structures, the onset of snap-through is governed by bifurcations whose symmetry and dynamics set the threshold and the direction of the jump~\cite{radisson2023elastic, radisson2023dynamic, wang2024transient,gomez2017critical, pandey2014dynamics}. Near instability, rate effects such as delayed bifurcation shift the critical displacement through inertia or material dissipation~\cite{urbach2020predicting, sano2018snap, liu2021delayed}. Viscoelastic structures can also snap after a delay under a fixed load, as the material relaxes~\cite{gomez2019dynamics,steinmontalvo2021delayed}. Such delays have been programmed to last from seconds to days~\cite{qing2025programmable}. Incorporating viscoelastic snapping elements into architected materials has opened a route to rate-programmable mechanical function~\cite{dykstra2019viscoelastic,janbaz2020strain,dykstra2022extreme}.

These rate effects, however, all originate in the bulk. When elastic elements instead interact through contact, the instability is no longer set by the shape of each element alone.  Multiple contact points reshape the energy landscape of buckled beams, enabling multi-step snapping and geometric control of instability onset~\cite{meulblok2025early,kwakernaak2024collective,ma2023contact,dura2026hysteretic}. Similarly, closely spaced beam arrays exhibit ordered configurations and collective instabilities mediated by contact constraints~\cite{guerra2023self,virgin2018tailored,virgin2021simultaneous}. These advances have focused on quasistatic behavior, in which contact modifies where snapping occurs but not how it unfolds in time~\cite{kwakernaak2024collective}. Yet frictional and adhesive interfaces are inherently time dependent. Held below the load at which they slide, frictional interfaces creep. This creep slows as the contact ages and strengthens, so that a stationary interface heals rather than fails~\cite{baumberger2006solid,sirorattanakul2025sliding}; close to the sliding threshold, creep either jams or accelerates into sliding, depending on the load~\cite{bureau2001jamming}. Adhesive contacts, in contrast, can detach suddenly after prolonged adherence, with a lifetime set by the competition between bulk rheology and interfacial viscoelasticity~\cite{linghu2025long}. Elastomer contacts also lose real contact area under increasing shear, well before they slide~\cite{sahli2018evolution}. Whether such interfacial kinetics can govern an elastic instability, setting when, where, and whether a structure snaps, has not been explored.

In this work, we show that contact between two buckled beams reshapes the dynamic pathway to snap-through under cyclic compression. Depending on the imposed compression rate, the beam pair exhibits one of three regimes: snap on loading (SoL), snap on unloading (SoU), or a packed state without snapping within the cycle. The loading and unloading branches follow distinct trajectories of rate-dependent delay, and the unloading branch carries a memory of the loading rate. We trace this delay to a frictional rolling-to-sliding transition at the beam interface. Faster driving leaves less time before sliding, carrying the snap to larger compression, and beyond a critical rate, out of the cycle altogether. When the compression is instead stopped beyond a threshold, the pair snaps after a wait that varies from seconds to hours. The held contact thus fails rather than heals. Exploiting this delay in arrays of beam pairs with graded gaps, we program rate-dependent sequences of snap events. The loading and unloading rates serve as two independent controls, as in dynamic driving of mechanical bits~\cite{gutierrezprieto2026dynamic}, but here the rate sensitivity comes from the contact. The energy dissipated per cycle then becomes a two-parameter function of the two rates, increasing fourfold when, at the fastest loading, the return is slowed.

\begin{figure*}[!tbp]
\centering
\includegraphics[width=.95\textwidth]{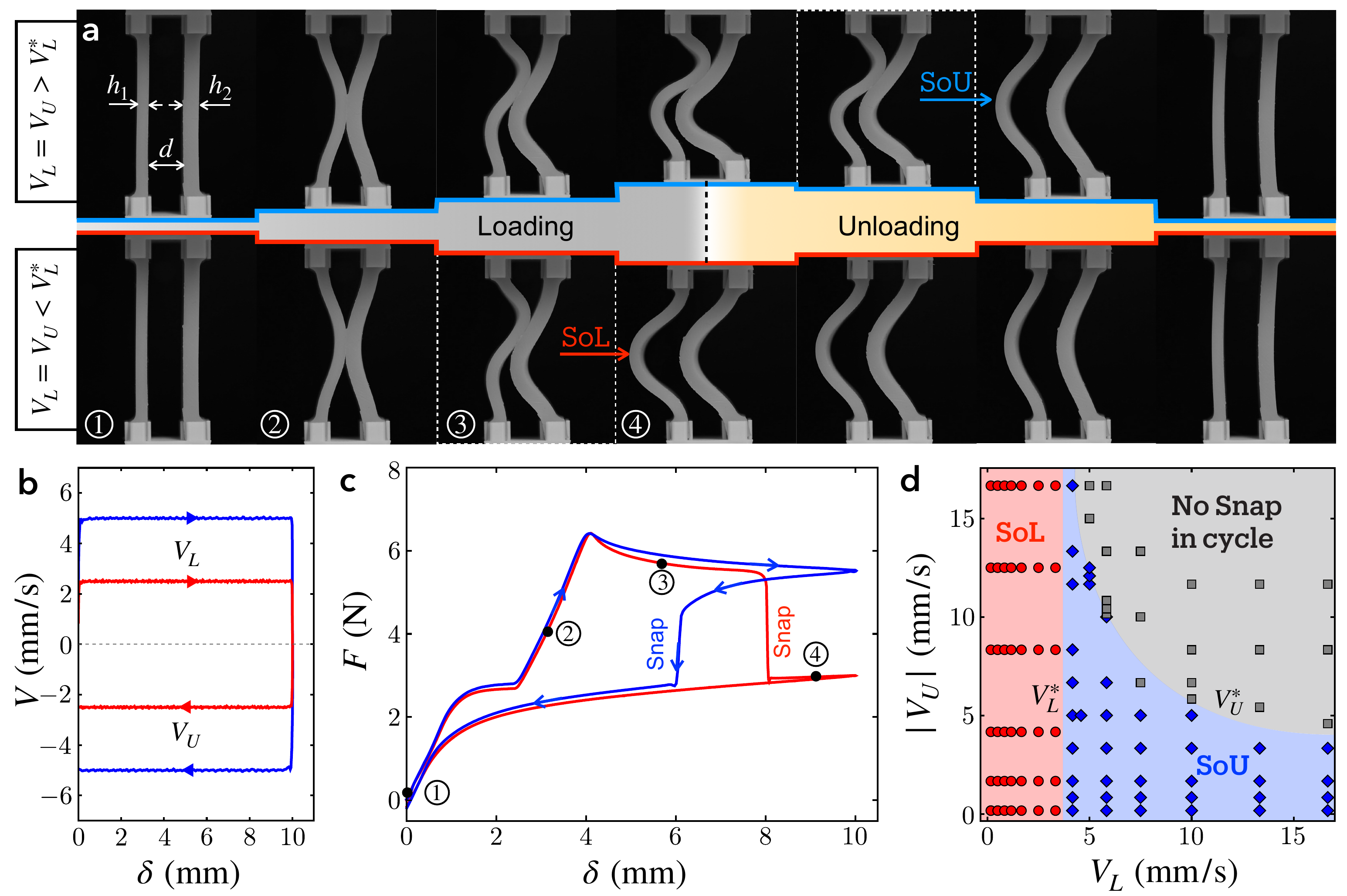}
\caption{\textbf{Rate dependent snap buckling during cyclic loading and return of contacting beams}. \textbf{a}. Snapshots of a pair of elastic beams of unequal thickness subjected to cyclic axial compression at two different rates, $V_L=V_U=5$~mm/s (upper row) and $V_L=V_U=2.5$~mm/s (lower row). At slower rates the pair snaps on the loading branch (snap on loading, SoL; lower row), whereas at higher rates snapping occurs on the unloading branch (snap on unloading, SoU; upper row). The maximum compression in both cases is $\delta_{\max} = 10$~mm, and the gap is $d = 10$~mm. \textbf{b}. Imposed displacement protocol showing piecewise constant loading and unloading speeds $V_L$ and $V_U$. \textbf{c}. Representative force--displacement curves for two rate protocols (red, SoL; blue, SoU). Snap-through is marked by a discontinuous drop in force. \textbf{d}. Regime map in the ($V_L$, $V_U$) plane showing snap on loading (SoL), snap on unloading (SoU), and packed configurations where snap through is suppressed, separated by the critical loading and unloading rates $V_L^\ast$ and $V_U^\ast$}.
\label{fig1}
\vspace{-7mm}
\end{figure*}

\section*{Results}

\subsection*{Experiments}
\begin{figure*}[!tbp]
\centering
\includegraphics[width=.9\textwidth]{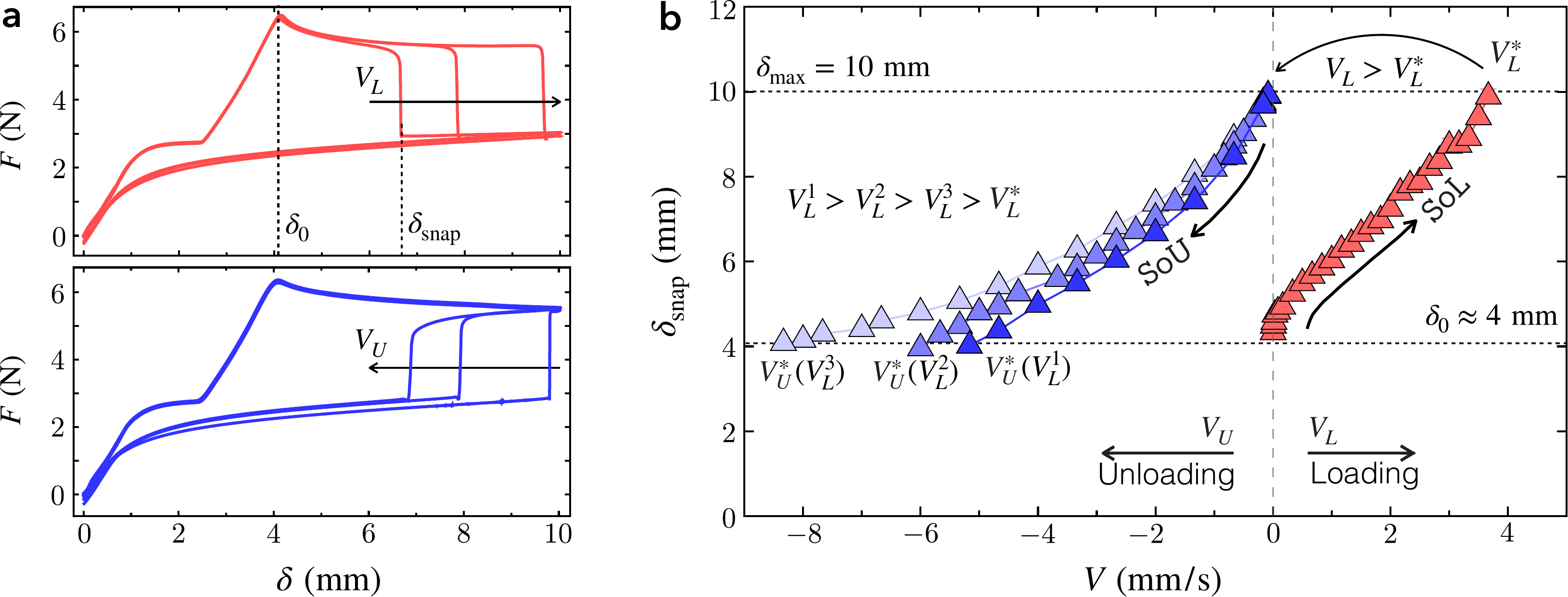}
\caption{\textbf{Asymmetric delay of snap-through on the loading and unloading branches}. a. Force--displacement curves at increasing speed for the loading (top, red) and unloading (bottom, blue) branches; the snap location $\delta_{\text{snap}}$ shifts with rate while the pre-snap curve is unchanged. b. Snap displacement $\delta_{\text{snap}}$ versus imposed rate $V$ (positive $V$: loading/SoL, negative $V$: unloading/SoU), with shading from light to dark in the SoU data indicating increasing $V_L$. The dashed line marks the characteristic displacement $\delta_0\approx4$~mm, the force peak and the onset of the mode-1 to mode-2 transition. On loading, $\delta_{\text{snap}}$ grows with rate until it reaches $\delta_{\max}$ at the critical loading rate $V_L^{\ast}$; on unloading, $\delta_{\text{snap}}$ decreases from $\delta_{\max}$ toward $\delta_0$ at the critical unloading rate $V_U^{\ast}$. The three unloading branches (blue data points) correspond to prior loading rates $V_L^1 (13.33\,\text{mm/s})> V_L^2 (8.33\,\text{mm/s}) > V_L^3 (5\,\text{mm/s}) > V_L^{\ast}$. }
\label{fig2}
\vspace{-3mm}
\end{figure*}

We compress a pair of clamped beams of length $44$~mm and unequal thicknesses $h_1 = 3$~mm and $h_2 = 4$~mm, separated by a gap $d$ (Fig.~\ref{fig1}a). Both beams are made of a silicone elastomer (vinylpolysiloxane, Zhermack Elite Double 32) with Young's modulus $E = 1.2$~MPa. The gap is varied between experiments and set to $d = 10$~mm throughout Figs.~\ref{fig1} and \ref{fig2}. The outer edges of the beams near the clamps are rounded to ensure that upon axial compression they buckle toward each other and establish contact (mode-1; Fig.~\ref{fig1}a, column 2)~\cite{PhysRevLett.130.268204}. As compression increases, the thinner beam transitions to a higher buckled configuration (mode-2; Fig.~\ref{fig1}a, column 3) while maintaining contact with the thicker beam. It then snaps laterally, at the snap displacement $\delta_{\text{snap}}$. The thickness difference and choice of $d$ ensure that the thicker beam consistently biases the thinner beam toward a single snapping direction~\cite{kwakernaak2024collective}. We subject the pair to cyclic compression with independently prescribed loading and unloading speeds $V_L$ and $V_U$. For the image sequence shown in Fig.~\ref{fig1}, we set $V_L=V_U$ and the maximum compression at $\delta_{\max} = 10$~mm, as shown in the prescribed loading-unloading cycle of Fig.~\ref{fig1}b. 

We observe three distinct responses, set by the imposed driving rate (Movie S1). At lower rates, snap-through occurs on the loading branch (SoL; $V_L=V_U=2.5$~mm/s in Fig.~\ref{fig1}a). At higher loading rates, the beams do not snap during loading; instead, snap-through occurs only on the return branch, when the pair is returned at the same speed (SoU; $5$~mm/s in Fig.~\ref{fig1}a). The force--displacement curves in Fig.~\ref{fig1}c reflect this branch selection. During loading, the force increases smoothly as the beams buckle and come into contact (state \textcircled{1}-\textcircled{2}), followed by a regime of contact-mediated deformation in which the thinner beam transitions toward a higher mode while maintaining contact (state \textcircled{3}). Snap-through corresponds to a discontinuous force drop (state \textcircled{3}-\textcircled{4}), marking rapid loss of contact. In the SoL regime, this discontinuity occurs on loading (red curve), whereas in the SoU regime the loading branch remains continuous up to $\delta_{\max}$, and the force drop appears only during return (blue curve). At the highest rates, the beams remain packed in contact throughout the cycle and do not snap within it; the force--displacement response stays continuous on both branches, with no discontinuous drop (Fig.~S2).

\begin{figure}[!tbp]
\centering
\includegraphics[width=.45\textwidth]{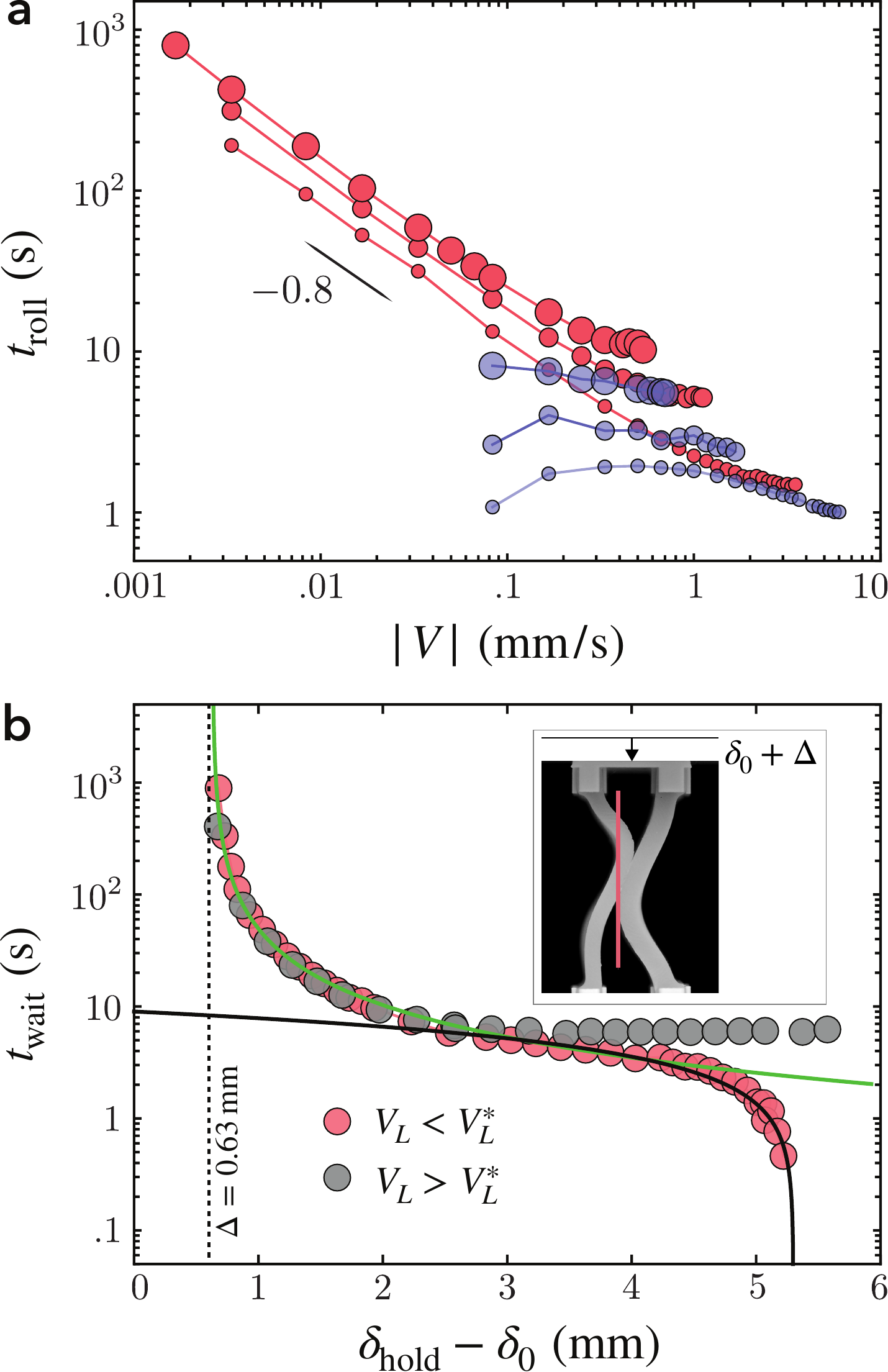}
\caption{\textbf{ Rolling-to-sliding transition under continuous driving and in hold experiments}. \textbf{(a)} Rolling time $t_{\mathrm{roll}}$, the interval the pair spends rolling before it slides and snaps, versus imposed speed $|V|$ for snap on loading (red) and snap on unloading (blue). Increasing marker sizes represent larger gaps ($d = 10$, $11$ and $12$~mm). The loading branch follows an empirical power law $t_{\mathrm{roll}}\sim|V|^{-0.8}$ at low speed (guide line), and each series terminates at its own critical loading rate. \textbf{(b)} Wait time $t_{\mathrm{wait}}$ after the compression is stopped at $\delta_{\mathrm{hold}}>\delta_0$, plotted against $\delta_{\mathrm{hold}}-\delta_0$, at $d = 12$~mm, for a loading rate below ($V_L = 0.5$~mm/s) and above ($V_L = 0.83$~mm/s) the critical rate $V_L^{\ast}\simeq0.7$~mm/s. The dotted line marks the offset $\Delta\simeq0.6$~mm at which $t_{\mathrm{wait}}$ diverges; the wait time vanishes for $V_L < V_L^{\ast}$ as $\delta_{\text{hold}}\rightarrow\delta_{\text{snap}}$ for the corresponding $V_L$. Solid lines represent power-law fits to the two limits, with exponents $-1.2$ (green line) and $0.65$ (black line). Inset shows the geometry of the beam pair at a compression of $\delta_0 + \Delta$.}
\label{fig4}
\vspace{-3mm}
\end{figure}

We vary $V_L$ and $V_U$ systematically, up to $17$~mm/s, and map three dynamical regimes in the $(V_L, V_U)$ plane (Fig.~\ref{fig1}d). This plane is the control space of the pair. Each protocol is a point in it, and two boundaries partition it into the three outcomes. The two rates select whether, and on which branch, the pair snaps. The boundary separating SoL from the remaining regimes is a critical loading rate $V_L^{\ast}\approx3.7$~mm/s. For $V_L < V_L^*$, the beams always lose contact during loading, and the subsequent unloading branch returns the snapped configuration to its initial state. For $V_L > V_L^*$, contact is maintained throughout loading, and any snap occurs on the return. In this regime, SoU is observed below a critical unloading rate $V_U^*$, above which the pair stays packed through the whole cycle. The two thresholds are different in character. The loading threshold $V_L^*$ is set on the loading branch alone, whereas $V_U^*$ depends on $V_L$. The beam pair thus arrives at maximum compression carrying a memory of the rate at which it was compressed. In the regime map, this memory is the curved boundary of the SoU region, along which $V_U^*$ falls as $V_L$ rises. We trace this history dependence to the delayed onset of sliding at the contact (Fig.~\ref{fig4} and figs.~S3 and S4) and harness it for programmable, history-dependent sequences (Fig.~\ref{fig5}).

\subsection*{Asymmetric delay on the loading and unloading branches}
To quantify the rate dependence of snap-through on each branch, we plot the force--displacement curves for different loading and unloading rates (Fig.~\ref{fig2}a).  Over more than an order of magnitude in rate, the curves overlap up to the discontinuous force drop that marks the snap. This invariance shows that the bulk response of the beams is rate independent over this range, and that the rate enters only through where the snap occurs. The peak-force point is at $\delta_0 \approx 4~\mathrm{mm}$, which coincides with the onset of the mode-1 to mode-2 transition of the thinner beam. Beyond $\delta_0$, the force--displacement curve exhibits a regime of negative incremental stiffness preceding the abrupt loss of contact. As the loading rate increases, the snap location moves to larger compressions and eventually approaches the imposed maximum compression $\delta_{\max}$. Above $V_L^{\ast}$, the snap moves onto the unloading branch. On unloading, the snap displacement decreases with increasing return rate (Fig.~\ref{fig2}a, bottom), so a faster return carries the rolling contact further back, closer to $\delta_0$, before sliding begins.

We extract $\delta_{\text{snap}}$ from the force--displacement curves (Fig.~\ref{fig2}a) and plot it against the imposed rate $V$, positive for loading and negative for unloading (Fig.~\ref{fig2}b). For the loading branch, $\delta_{\text{snap}}$ increases monotonically with $V_L$, moving toward $\delta_0$ as $V_L$ decreases, and reaching $\delta_{\max}$ at the critical loading rate $V_L = V_L^{\ast}$. Beyond $V_L^{\ast}$, the pair stays in contact throughout loading. On the unloading branch ($V<0$), snapping initiates from the maximally compressed state $\delta_{\max}$, and $\delta_{\text{snap}}$ decreases with increasing unloading rate. The unloading branch ends near the same characteristic displacement $\delta_0$ at a critical rate $V_U^*$, beyond which the pair stays packed through the return. Notably, the unloading curves in Fig.~\ref{fig2}b separate according to the prior loading rate, so the delay carries a memory of that rate. The unloading branch starts from the packed state reached at $\delta_{\max}$, so the two branches begin from different configurations.

\subsection*{ Time-dependent onset of sliding as the origin of the snap delay}
Upon contact, the thinner beam does not immediately slide past its neighbor. Instead, it rolls along the thicker beam while transitioning from the first to the second buckling mode, the contact point migrating continuously and without interfacial slip. Snap-through is thus not set by the elastic instability alone but by the activation of sliding at the interface. We follow this stage by tracking the two contact edges $y_c^t$ and $y_c^b$ against compression (Fig.~S3). The edges coincide while the pair remains symmetric and bifurcate once one beam begins to roll over the other. The bifurcation happens at $\delta_0$, to within the $\pm0.5$~mm resolution of the edge tracking. At different loading rates the edges trace the same curve, so the contact configuration is set by the compression alone. Rolling ends abruptly when the beams slide apart and snap (Movie S2), and the faster the drive, the further the contact rolls before it slides. This rate dependence is best quantified by the rolling time $t_{\mathrm{roll}}$, the interval the beams spend rolling before they slide apart, measured from $\delta_0$ for snap on loading and from $\delta_{\max}$ for snap on unloading (Fig.~\ref{fig4}a).   As the speed rises, the snap moves to larger compression, yet $t_{\mathrm{roll}}$ falls, so faster driving delays the snap in displacement but advances it in time. A sliding threshold set by the contact configuration would fix the snap displacement and make $t_{\mathrm{roll}}$ fall as $1/|V|$, while a fixed time delay would leave $t_{\mathrm{roll}}$ unchanged. On the loading branch at low speed, where $\delta_{\text{snap}}$ is close to $\delta_0$, $t_{\mathrm{roll}}$ falls between these two limits, following an empirical power law $t_{\mathrm{roll}}\sim|V|^{-0.8}$. The onset of sliding thus depends jointly on the compression and on time. Pairs driven at different rates slide at different cofigurations. These data point to an internal variable of the contact that evolves in time while the contact is loaded and triggers sliding once it has advanced far enough. On the unloading branch, $t_{\mathrm{roll}}$ tends toward a weaker power-law decrease than on loading as faster return carries $\delta_{\text{snap}}$ toward $\delta_0$. The two branches differ both in where the snap occurs and in how the time to slide responds to speed.

Under continuous driving, compression and time advance together, so the rolling time cannot separate the role of the configuration from that of time. To separate the two, we stop the compression at a displacement $\delta_{\mathrm{hold}}>\delta_0$ and hold it there. If sliding were set by the configuration alone, a pair held below its snap displacement would never snap. If instead sliding is set by an internal variable that evolves in time, the held pair snaps after a delay. Measuring this delay at different hold displacements then shows how the variable responds to load. These experiments use two loading rates, one below and one above $V_L^{\ast}$. Once the compression is held, rolling ceases, and the pair snaps after a wait time $t_{\mathrm{wait}}$ that ranges from seconds to hours (Fig.~\ref{fig4}b). During the wait the force relaxes by a few percent (fig.~S4a), so the pair slides under a smaller load than it carried when the compression stopped. This points to a contact whose resistance falls with time. The contact edges stay fixed during the hold, to within the tracking resolution, so the internal variable advances at fixed configuration. Holds above $V_L^{\ast}$ also end in a snap, so the packed regime reflects the same delay, extended beyond the end of the cycle. The drive rate thus sets how far this variable advances within the cycle, and the maximum compression sets whether it advances far enough.

One might expect that once the beams are past the symmetric configuration and the force maximum at $\delta_0$, they will snap eventually. Instead, the wait time diverges at a finite displacement $\Delta$ beyond $\delta_0$, and a pair held within $\Delta\simeq0.6$~mm of $\delta_0$ does not snap over the hours we observe. The finite $\Delta$ shows that the internal variable advances toward sliding only above a minimum load. It comes from the finite thickness of the beams and corresponds to the compression at which the left edge of the left beam aligns with the left edge of the right beam in the twisted configuration (Fig.~\ref{fig4}b, inset). Below this compression, the left beam has not been displaced far enough to the right to slide off its neighbour, however long it is held. At the other end, for $V_L < V_L^{\ast}$, the wait time vanishes as $\delta_{\mathrm{hold}}$ approaches $\delta_{\text{snap}}(V_L)$, the snap displacement of the continuous run at the same rate. Between these two limits the wait falls steadily as $\delta_{\mathrm{hold}}$ rises, so the variable advances faster the further the contact is loaded. The solid curves in Fig.~\ref{fig4}b are empirical power laws in the distance to each limit. Near the divergence,
\begin{equation}
t_{\mathrm{wait}} \sim \left(\delta_{\mathrm{hold}}-\delta_0-\Delta\right)^{-1.2},
\label{eq:diverge}
\end{equation}
over two decades in $\delta_{\mathrm{hold}}-\delta_0-\Delta$, and near $\delta_{\text{snap}}$,
\begin{equation}
t_{\mathrm{wait}} \sim \left(\delta_{\text{snap}}-\delta_{\mathrm{hold}}\right)^{0.65}
\label{eq:approach}
\end{equation}
over more than a decade (Fig.~S4, b and c).

The two datasets in Fig.~\ref{fig4}b, despite their different loading rates, overlap in the region close to $\delta_0$. This is where the wait time diverges, reaching hundreds of seconds, far longer than the compression itself lasts. Far from $\delta_0$ the waits fall to a few seconds and the datasets separate, the pair loaded more slowly snapping sooner, while the faster run flattens onto a plateau near six seconds. 
The two curves deviate from each other when the wait time decreases and becomes comparable to the time spent by the beams under compressive driving.
No trigger set by the configuration alone can show this dependence. For loading rates above $V_L^{\ast}$, a slower compression spends longer past the force peak, where the internal variable advances, so the pair arrives at $\delta_{\max}$ with the variable closer to the sliding onset. The return then needs less time to complete the advance, and the pair snaps even when it is unloaded quickly, which raises $V_U^{\ast}$. This effect is also seen in Figs.~\ref{fig1}d and \ref{fig2}b, where $V_U^{\ast}$ falls as $V_L$ rises. This effect is also seen in Figs.~\ref{fig1}d and \ref{fig2}b, where $V_U^{\ast}$ falls as $V_L$ rises. Because rolling is slip-free, the contact configuration at $\delta_{\max}$ is the same for every loading rate, so this memory of the loading rate is carried by the internal variable, not by the geometry of the contact. Sliding at this contact is thus delayed, beginning after a time set by the compression and its history. This delayed sliding is analogous to delayed yielding in soft glassy materials and creep rupture in fibre bundles, where failure follows a delay above a threshold~\cite{bonn2017yield,kun2003creep}. These measurements establish that the internal variable exists and how it behaves, but not what it is; identifying it, and explaining the exponents in Eqs.~\eqref{eq:diverge} and \eqref{eq:approach}, requires further theoretical, numerical and experimental work.

\begin{figure*}[!tbp]
\centering
\includegraphics[width=1\textwidth]{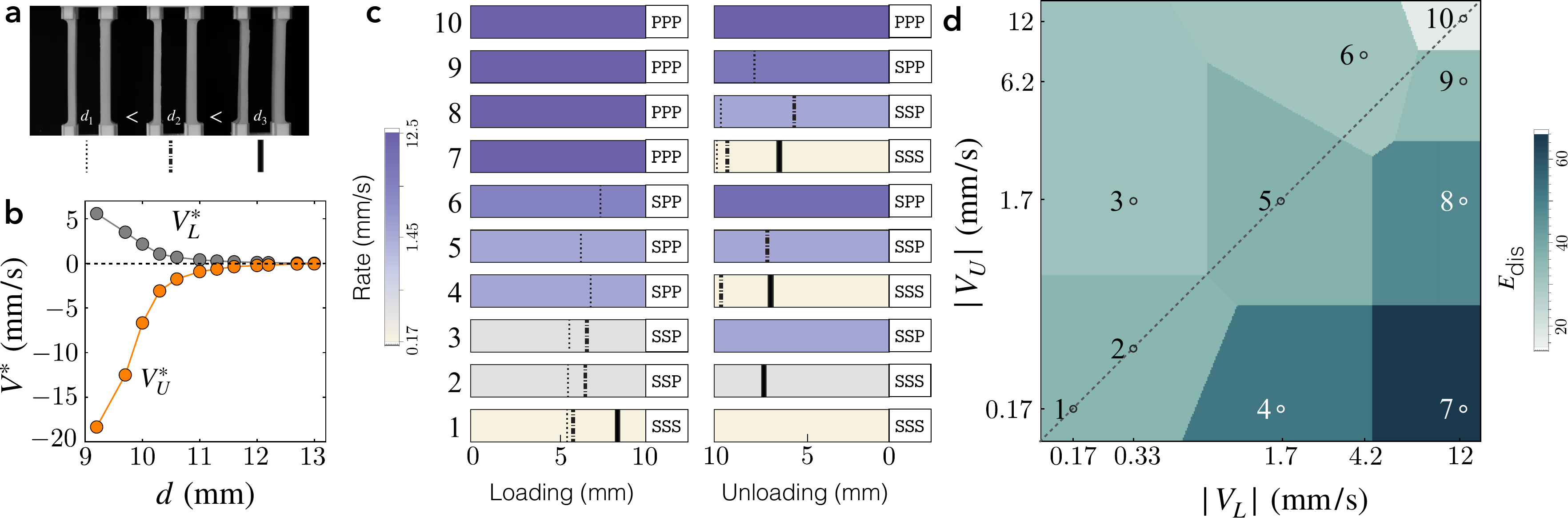}
\caption{\textbf{Tunable sequence and energetics of snapping in a gap-graded beam-pair array.}
\textbf{(a)} Three beam pairs with increasing gaps, $d_1 < d_2 < d_3$; the line styles below encode each gap and mark the snaps in (c).
\textbf{(b)} Critical rates against gap: $V_L^{\ast}$ (gray) and $V_U^{\ast}$ (orange, negative for the return), both falling toward zero as $d$ increases. \textbf{(c)} Snap ``score'' for ten loading--unloading protocols (rows). Each row shows the loading branch (left, $\delta: 0 \to 10$~mm) and the unloading branch (right, $\delta: 10 \to 0$~mm) as bars colored by the imposed rate ($V_L$ and $V_U$; colorbar). Vertical lines mark where each pair snaps, on the branch where it occurs, at its snap displacement, with thickness set by that pair's gap. Right-hand labels give the cumulative configuration of the three pairs at the end of each branch (S, snapped; P, packed). From fast and fully packed (top) to slow and fully snap-on-loading (bottom), the rate protocol selects distinct snap sequences and final states.
\textbf{(d)} Energy dissipated per cycle, $E_{\mathrm{dis}}$, for the ten protocols of (c), placed at their rates $(|V_L|, |V_U|)$ (circles). Each protocol gives one value, and the shading is a Voronoi tessellation of those points within a cell representing the value of $E_{\mathrm{dis}}$. }
\label{fig5}
\vspace{-3mm}
\end{figure*}

\subsection*{Rate-programmed snap sequences and dissipation in gap-graded arrays}
Having established how the snap is delayed in a single pair, we now use that delay as a design tool. A single pair converts the imposed rate into one of three outcomes: snap on loading, snap on unloading, or packed, the three regimes of Fig.~\ref{fig1}d. In an array of pairs with graded gaps, the two rates of one global drive then program an ordered response. We demonstrate this with three beam pairs placed side by side, each with a different gap ($d_1 < d_2 < d_3$, Fig.~\ref{fig5}a). The pairs share the same imposed displacement and rate, and are mechanically independent. Because the gap sets each pair's critical rates, the pairs change outcome in a fixed order. The state of the array is then specified by the number of pairs in each outcome: for $N$ pairs, $(N+1)(N+2)/2$ states, ten for the three pairs used here. We reach these states by tuning the loading and unloading rates (Fig.~\ref{fig5}c and Movie S3).
At the lowest rates, with loading and unloading equal, all three pairs snap on loading. Increasing the rate moves the snaps onto the unloading branch and then suppresses them entirely, with the largest-gap pair leaving the loading branch first. With unequal rates, the loading rate alone sets how many pairs snap on loading. How many of the remaining pairs snap on unloading depends on both rates, since the pairs reach $\delta_{\max}$ carrying a memory of the loading rate. Sweeping the protocol this way realizes all ten states, so a single global drive encodes an ordered, history-dependent sequence and records how the array was driven and persists until the compression is fully released. The array thus acts as a set of independent relay hysterons with ordered thresholds, driven by two rates rather than one amplitude~\cite{preisach1935magnetische,keim2019memory}.

The branch asymmetry has a direct energetic consequence. The snap delay differs on the two branches and the two rates are set independently, so the dissipation depends on both rates. The energy dissipated per cycle, $E_{\mathrm{dis}}$, is the area enclosed by the array's force--displacement loop. Figure~\ref{fig5}d shows it against the two rates for the ten protocols of Fig.~\ref{fig5}c, and Fig.~S5 shows corresponding force--displacement cycles with the enclosed area shaded. Because the geometry and the elastic material are fixed, the rates alone select the dissipation. Along the diagonal $V_L = V_U$, $E_{\mathrm{dis}}$ stays nearly constant over a decade in rate while the pairs snap. It falls by more than half once all three pairs stay packed. The snaps carry most of the dissipation, and the rate acts through whether and when they occur. $E_{\mathrm{dis}}$ is asymmetric about the diagonal and is largest when the array is loaded quickly and unloaded slowly. Exchanging the two rates alters the dissipated energy (protocols 3 and 4) because it shifts when the pairs snap. A slow loading makes them snap early, at small compression, whereas a fast loading packs them and a slow return lets them snap from the packed state. With all three pairs snapping in both cases, snapping on the return dissipates nearly twice as much as snapping on loading (protocols 7 and 1). At the fastest loading, the return rate alone decides the outcome. A fast return leaves the three pairs packed until the load is removed, whereas a slow return lets all three snap and dissipates about four times more. The energy stored in the packed array can thus be recovered or released by choosing the return rate after the load is applied.

\section*{Discussion}
In summary, we have shown that contact makes snap-through a far richer transition than it is in an isolated beam. The same cyclic drive produces three distinct outcomes, selected by rate alone. Where the snap occurs depends jointly on the configuration of the contact, on the rate at which it is approached, and on the history of the drive. The unloading branch retains a memory of how fast the pair was compressed. The interface, rather than the energy landscape, sets the transition. Sliding begins after a time set by the compression and its history, and the drive determines on which branch, if any, that time is reached. These results imply that wherever slender structures are densely packed and interacting, their instabilities may be governed as much by the time dependence of their contacts as by their shapes.

In flexible metamaterials, rate sensitivity has been built in through viscoelastic constituents~\cite{dykstra2019viscoelastic,janbaz2020strain,dykstra2022extreme}, and dissipation is controlled by the geometry of snapping elements~\cite{shan2015multistable,restrepo2015phase,frenzel2016tailored}, reprogrammed by changing their confinement~\cite{almessabi2024reprogramming}, or tuned by external fields~\cite{spencer1997phenomenological}. Here the rate sensitivity comes from the contact in an elastic material, and the order of the two rates selects the dissipation. The energy stored in a packed array can then be released or recovered by choosing the return rate after the load is applied. Arrays with more pairs, graded in gap and thickness, could tailor a richer dissipation landscape, toward passive absorbers tuned by how fast a load is withdrawn. The contact itself, through its roughness, lubrication or surface chemistry, could become a design parameter alongside geometry and material~\cite{sahli2024frictional}. With suitable modifications, assemblies of such contact-mediated elements could also show a structural Mullins effect, a softening that records the largest deformation previously applied~\cite{keim2019memory,diani2009review}. Under repeated cycles, the pattern of snapped elements would then record the maximum compression of the previous cycle. More broadly, the function here comes not from permanent connections between elements but from contacts that form and break during the drive. Contact-mediated interactions of this kind could build timing, memory and rate-programmable energy absorption into structures that carry no circuitry of their own.

\section*{Materials and Methods}
\subsection*{Sample fabrication} We mixed the two components of a vinylpolysiloxane elastomer (Zhermack Elite Double 32) at $1\!:\!1$ by weight and cast the mixture, without degassing, in a 3D printed mold (Fig.~S1, inset). The mold set the beam length, $44$~mm, the two thicknesses, $h_1 = 3$~mm and $h_2 = 4$~mm, and the rounded outer edges near the ends. Each beam stayed in the mold for seven days at room temperature, so that the cure ran to completion and the beam was released straight and free of residual stress. A built-in curvature would otherwise add a second bias to the snapping direction, set by the casting rather than by the thickness difference. The cured material has Young's modulus $E \approx 1.2$~MPa~\cite{gutierrezprieto2026dynamic}.

\subsection*{Mechanical testing and data analysis}
Each beam was held at both ends in custom clamps that fix the clamped length and hold the two beams parallel at a gap $d$ (Fig.~S1), and we changed $d$ by exchanging the clamps. The clamps were attached to a compression plate mounted in a universal testing machine (Instron 34SC-1) with a $100$~N load cell. All experiments ran under displacement control at room temperature: the machine imposed the axial displacement $\delta$ at a prescribed rate and recorded the force $F$, while a camera (Nikon Z5) imaged the pair. A cycle is a compression to $\delta_{\max} = 10$~mm followed by a return to $\delta = 0$, at independently prescribed loading and unloading speeds $V_L$ and $V_U$. For the array, the three pairs were placed side by side and driven by the same crosshead, so that the load cell recorded their total force. In the hold experiments, at $d = 12$~mm, the crosshead stopped at $\delta_{\mathrm{hold}}$ and held that position until the pair snapped or the observation was ended, and $t_{\mathrm{wait}}$ is measured from the moment the crosshead stops. The snap is identified by the discontinuous drop in $F$, which gives $\delta_{\text{snap}}$, and $t_{\mathrm{roll}}$ is measured to the snap from $\delta_0$ on loading and from $\delta_{\max}$ on unloading. The contact edges (Fig.~S3) are tracked from the images with a resolution of $\pm0.5$~mm. The energy dissipated per cycle, $E_{\mathrm{dis}}$, is the area enclosed by the force--displacement cycle of the array. Each protocol was repeated 3-5 times.

\section*{Acknowledgements}
We acknowledge helpful discussions with Chris Santangelo and Dominic Vella. G.L. acknowledges funding from the Syracuse Office of Undergraduate Research \& Creative Engagement (The SOURCE). S.M. acknowledges funding from University Fellowship.

\bibliographystyle{apsrev4-2}
 \bibliography{snap}

@article{rafsanjani2015snapping,
  title={Snapping mechanical metamaterials under tension},
  author={Rafsanjani, Ahmad and Akbarzadeh, Abdolhamid and Pasini, Damiano},
  journal={Advanced Materials},
  volume={27},
  number={39},
  pages={5931--5935},
  year={2015},
  doi={10.1002/adma.201502809},
  publisher={Wiley}
}

@article{ma2023contact,
  title={Contact-driven snapping in thermally actuated metamaterials for fully reversible functionality},
  author={Ma, Ruizhe and Wu, Lei and Pasini, Damiano},
  journal={Advanced Functional Materials},
  volume={33},
  number={16},
  pages={2213371},
  year={2023},
  doi={10.1002/adfm.202213371},
  publisher={Wiley}
}

@article{almessabi2024reprogramming,
  title={Reprogramming multi-stable snapping and energy dissipation in origami metamaterials through panel confinement},
  author={Almessabi, Abdulrahman and Li, Xuwen and Jamalimehr, Amin and Pasini, Damiano},
  journal={Philosophical Transactions of the Royal Society A},
  volume={382},
  number={2283},
  pages={20240005},
  year={2024},
  doi={10.1098/rsta.2024.0005},
  publisher={The Royal Society}
}

@article{PhysRevLett.130.268204,
  title = {Counting and Sequential Information Processing in Mechanical Metamaterials},
  author = {Kwakernaak, Lennard J. and van Hecke, Martin},
  journal = {Phys. Rev. Lett.},
  volume = {130},
  issue = {26},
  pages = {268204},
  numpages = {5},
  year = {2023},
  month = {Jun},
  publisher = {American Physical Society},
  doi = {10.1103/PhysRevLett.130.268204},
  url = {https://link.aps.org/doi/10.1103/PhysRevLett.130.268204}
}

@article{wang2024transient,
  title={Transient amplification of broken symmetry in elastic snap-through},
  author={Wang, Qiong and Giudici, Andrea and Huang, Weicheng and Wang, Yuzhe and Liu, Mingchao and Tawfick, Sameh and Vella, Dominic},
  journal={Physical Review Letters},
  volume={132},
  number={26},
  pages={267201},
  year={2024},
  publisher={APS}
}

@article{liu2021delayed,
  title={Delayed bifurcation in elastic snap-through instabilities},
  author={Liu, Mingchao and Gomez, Michael and Vella, Dominic},
  journal={Journal of the Mechanics and Physics of Solids},
  volume={151},
  pages={104386},
  year={2021},
  publisher={Elsevier}
}

@article{radisson2023elastic,
  title={Elastic snap-through instabilities are governed by geometric symmetries},
  author={Radisson, Basile and Kanso, Eva},
  journal={Physical Review Letters},
  volume={130},
  number={23},
  pages={236102},
  year={2023},
  publisher={APS}
}

@article{radisson2023dynamic,
  title={Dynamic behavior of elastic strips near shape transitions},
  author={Radisson, Basile and Kanso, Eva},
  journal={Physical Review E},
  volume={107},
  number={6},
  pages={065001},
  year={2023},
  publisher={APS}
}

@article{wang2023insect,
  title={Insect-scale jumping robots enabled by a dynamic buckling cascade},
  author={Wang, Yuzhe and Wang, Qiong and Liu, Mingchao and Qin, Yimeng and Cheng, Liuyang and Bolmin, Ophelia and Alleyne, Marianne and Wissa, Aimy and Baughman, Ray H and Vella, Dominic and others},
  journal={Proceedings of the National Academy of Sciences},
  volume={120},
  number={5},
  pages={e2210651120},
  year={2023},
  publisher={National Academy of Sciences}
}

@article{gomez2017critical,
  title={Critical slowing down in purely elastic ?snap-through? instabilities},
  author={Gomez, Michael and Moulton, Derek E and Vella, Dominic},
  journal={Nature Physics},
  volume={13},
  number={2},
  pages={142--145},
  year={2017},
  publisher={Nature Publishing Group UK London}
}

@article{kwakernaak2024collective,
  title={The collective snapping of a pair of bumping buckled beams},
  author={Kwakernaak, Lennard J and Guerra, Arman and Holmes, Douglas P and van Hecke, Martin},
  journal={Extreme Mechanics Letters},
  volume={69},
  pages={102160},
  year={2024},
  publisher={Elsevier}
}

@article{guerra2023self,
  title={Self-ordering of buckling, bending, and bumping beams},
  author={Guerra, Arman and Slim, Anja C and Holmes, Douglas P and Kodio, Ousmane},
  journal={Physical Review Letters},
  volume={130},
  number={14},
  pages={148201},
  year={2023},
  publisher={APS}
}

@article{vella2019buffering,
  title={Buffering by buckling as a route for elastic deformation},
  author={Vella, Dominic},
  journal={Nature Reviews Physics},
  volume={1},
  number={7},
  pages={425--436},
  year={2019},
  publisher={Nature Publishing Group UK London}
}

@article{gomez2017passive,
  title={Passive control of viscous flow via elastic snap-through},
  author={Gomez, Michael and Moulton, Derek E and Vella, Dominic},
  journal={Physical Review Letters},
  volume={119},
  number={14},
  pages={144502},
  year={2017},
  publisher={APS}
}

@article{meulblok2025early,
  title={Early onset of snapping of slender beams under transverse forcing},
  author={Meulblok, Colin M and Bense, Hadrien and Caelen, M and van Hecke, Martin},
  journal={Extreme Mechanics Letters},
  volume={80},
  pages={102407},
  year={2025},
  publisher={Elsevier}
}

@article{drotman2021electronics,
  title={Electronics-free pneumatic circuits for controlling soft-legged robots},
  author={Drotman, Dylan and Jadhav, Saurabh and Sharp, David and Chan, Christian and Tolley, Michael T},
  journal={Science Robotics},
  volume={6},
  number={51},
  pages={eaay2627},
  year={2021},
  publisher={American Association for the Advancement of Science}
}

@article{rothemund2018soft,
  title={A soft, bistable valve for autonomous control of soft actuators},
  author={Rothemund, Philipp and Ainla, Alar and Belding, Lee and Preston, Daniel J and Kurihara, Sarah and Suo, Zhigang and Whitesides, George M},
  journal={Science Robotics},
  volume={3},
  number={16},
  pages={eaar7986},
  year={2018},
  publisher={American Association for the Advancement of Science}
}

@article{pandey2014dynamics,
  title={Dynamics of snapping beams and jumping poppers},
  author={Pandey, Anupam and Moulton, Derek E and Vella, Dominic and Holmes, Douglas P},
  journal={Europhysics Letters},
  volume={105},
  number={2},
  pages={24001},
  year={2014},
  publisher={EDP Sciences, IOP Publishing and Societ{\`a} Italiana di Fisica}
}

@article{urbach2020predicting,
  title={Predicting delayed instabilities in viscoelastic solids},
  author={Urbach, Erez Y and Efrati, Efi},
  journal={Science Advances},
  volume={6},
  number={36},
  pages={eabb2948},
  year={2020},
  publisher={American Association for the Advancement of Science}
}

@article{sano2018snap,
  title={Snap-buckling in asymmetrically constrained elastic strips},
  author={Sano, Tomohiko G and Wada, Hirofumi},
  journal={Physical Review E},
  volume={97},
  number={1},
  pages={013002},
  year={2018},
  publisher={APS}
}

@article{virgin2018tailored,
  title={Tailored buckling constrained by adjacent members},
  author={Virgin, Lawrence},
  journal={Structures},
  volume={16},
  pages={20--26},
  year={2018},
  publisher={Elsevier}
}

@article{virgin2021simultaneous,
  title={Simultaneous buckling, contact, and load-carrying capacity},
  author={Virgin, Lawrence},
  journal={Journal of Engineering Mechanics},
  volume={147},
  number={5},
  pages={04021023},
  year={2021},
  publisher={American Society of Civil Engineers}
}

@article{dura2026hysteretic,
  title={Hysteretic slit-snapping and multistability in buckled beams with partial cuts},
  author={Dur{\`a} Faul{\'\i}, Bernat and Kwakernaak, Lennard J and van Hecke, Martin},
  journal={Science Advances},
  volume={12},
  number={7},
  pages={eaeb9750},
  year={2026},
  publisher={American Association for the Advancement of Science}
}

@article{baumberger2006solid,
  title={Solid friction from stick--slip down to pinning and aging},
  author={Baumberger, Tristan and Caroli, Christiane},
  journal={Advances in Physics},
  volume={55},
  number={3-4},
  pages={279--348},
  year={2006},
  publisher={Taylor \& Francis}
}

@article{sahli2018evolution,
  title={Evolution of real contact area under shear and the value of static friction of soft materials},
  author={Sahli, R. and Pallares, G. and Ducottet, C. and Ben Ali, I. E. and Al Akhrass, S. and Guibert, M. and Scheibert, J.},
  journal={Proceedings of the National Academy of Sciences},
  volume={115},
  number={3},
  pages={471--476},
  year={2018},
  publisher={National Academy of Sciences}
}

@article{preisach1935magnetische,
  title={{\"U}ber die magnetische Nachwirkung},
  author={Preisach, F.},
  journal={Zeitschrift f{\"u}r Physik},
  volume={94},
  number={5-6},
  pages={277--302},
  year={1935},
  publisher={Springer}
}

@article{keim2019memory,
  title={Memory formation in matter},
  author={Keim, Nathan C. and Paulsen, Joseph D. and Zeravcic, Zorana and Sastry, Srikanth and Nagel, Sidney R.},
  journal={Reviews of Modern Physics},
  volume={91},
  number={3},
  pages={035002},
  year={2019},
  publisher={American Physical Society}
}

@article{gutierrezprieto2026dynamic,
  title={Dynamic drives allow independent control of material bits for targeted memory},
  author={Gutierrez-Prieto, Eduardo and Meulblok, Colin M and van Hecke, Martin and Reis, Pedro M},
  journal={Science Advances},
  volume={12},
  number={19},
  pages={eaec1606},
  year={2026},
  publisher={American Association for the Advancement of Science}
}

@article{sahli2024frictional,
  title={Frictional contact of soft polymeric shells},
  author={Sahli, Riad and Mikkelsen, Jeppe and Boye, Mathias S{\"a}therstr{\"o}m and Dias, Marcelo A and Aghababaei, Ramin},
  journal={Physical Review Letters},
  volume={133},
  number={10},
  pages={106202},
  year={2024},
  publisher={American Physical Society}
}

@article{sirorattanakul2025sliding,
  title={Sliding and healing of frictional interfaces that appear stationary},
  author={Sirorattanakul, Krittanon and Larochelle, Stacy and Rubino, Vito and Lapusta, Nadia and Rosakis, Ares J},
  journal={Nature},
  volume={639},
  number={8056},
  pages={947--953},
  year={2025},
  publisher={Springer Nature}
}

@article{bureau2001jamming,
  title={Jamming creep of a frictional interface},
  author={Bureau, L and Baumberger, T and Caroli, C},
  journal={Physical Review E},
  volume={64},
  number={3},
  pages={031502},
  year={2001},
  publisher={American Physical Society}
}

@article{gomez2019dynamics,
  title={Dynamics of viscoelastic snap-through},
  author={Gomez, Michael and Moulton, Derek E and Vella, Dominic},
  journal={Journal of the Mechanics and Physics of Solids},
  volume={124},
  pages={781--813},
  year={2019},
  publisher={Elsevier}
}

@article{steinmontalvo2021delayed,
  title={Delayed buckling of spherical shells due to viscoelastic knockdown of the critical load},
  author={Stein-Montalvo, Lucia and Holmes, Douglas P and Coupier, Gwennou},
  journal={Proceedings of the Royal Society A},
  volume={477},
  number={2253},
  pages={20210253},
  year={2021},
  publisher={The Royal Society}
}

@article{dykstra2019viscoelastic,
  title={Viscoelastic snapping metamaterials},
  author={Dykstra, David M J and Busink, Joris and Ennis, Bernard and Coulais, Corentin},
  journal={Journal of Applied Mechanics},
  volume={86},
  number={11},
  pages={111012},
  year={2019},
  publisher={ASME}
}

@article{dykstra2022extreme,
  title={The extreme mechanics of viscoelastic metamaterials},
  author={Dykstra, David M J and Janbaz, Shahram and Coulais, Corentin},
  journal={APL Materials},
  volume={10},
  number={8},
  pages={080702},
  year={2022},
  publisher={AIP Publishing}
}

@article{janbaz2020strain,
  title={Strain rate--dependent mechanical metamaterials},
  author={Janbaz, S and Narooei, K and van Manen, T and Zadpoor, A A},
  journal={Science Advances},
  volume={6},
  number={25},
  pages={eaba0616},
  year={2020},
  publisher={American Association for the Advancement of Science}
}

@article{qing2025programmable,
  title={Programmable seconds-to-days-long delayed snapping in jumping metashells},
  author={Qing, Haitao and Zhou, Caizhi and Qi, Fangjie and Yin, Jie},
  journal={Proceedings of the National Academy of Sciences},
  volume={122},
  number={23},
  pages={e2503313122},
  year={2025},
  publisher={National Academy of Sciences}
}

@article{linghu2025long,
  title={Long-term adhesion durability revealed through a rheological paradigm},
  author={Linghu, Changhong and Wu, Rui and Chen, Yuqing and Huang, Yulin and Seo, Young-Jae and Li, Hua and Wang, Guannan and Gao, Huajian and Hsia, K Jimmy},
  journal={Science Advances},
  volume={11},
  number={11},
  pages={eadt3957},
  year={2025},
  publisher={American Association for the Advancement of Science}
}

@article{bonn2017yield,
  title={Yield stress materials in soft condensed matter},
  author={Bonn, Daniel and Denn, Morton M and Berthier, Ludovic and Divoux, Thibaut and Manneville, S{\'e}bastien},
  journal={Reviews of Modern Physics},
  volume={89},
  number={3},
  pages={035005},
  year={2017},
  publisher={American Physical Society}
}

@article{kun2003creep,
  title={Creep rupture has two universality classes},
  author={Kun, F and Moreno, Y and Hidalgo, R C and Herrmann, H J},
  journal={Europhysics Letters},
  volume={63},
  number={3},
  pages={347--353},
  year={2003},
  publisher={IOP Publishing}
}

@article{shan2015multistable,
  title={Multistable architected materials for trapping elastic strain energy},
  author={Shan, Sicong and Kang, Sung H and Raney, Jordan R and Wang, Pai and Fang, Lichen and Candido, Francisco and Lewis, Jennifer A and Bertoldi, Katia},
  journal={Advanced Materials},
  volume={27},
  number={29},
  pages={4296--4301},
  year={2015},
  publisher={Wiley}
}

@article{restrepo2015phase,
  title={Phase transforming cellular materials},
  author={Restrepo, David and Mankame, Nilesh D and Zavattieri, Pablo D},
  journal={Extreme Mechanics Letters},
  volume={4},
  pages={52--60},
  year={2015},
  publisher={Elsevier}
}

@article{frenzel2016tailored,
  title={Tailored buckling microlattices as reusable light-weight shock absorbers},
  author={Frenzel, Tobias and Findeisen, Claudio and Kadic, Muamer and Gumbsch, Peter and Wegener, Martin},
  journal={Advanced Materials},
  volume={28},
  number={28},
  pages={5865--5870},
  year={2016},
  publisher={Wiley}
}

@article{spencer1997phenomenological,
  title={Phenomenological model for magnetorheological dampers},
  author={Spencer, B F and Dyke, S J and Sain, M K and Carlson, J D},
  journal={Journal of Engineering Mechanics},
  volume={123},
  number={3},
  pages={230--238},
  year={1997},
  publisher={American Society of Civil Engineers}
}

@article{diani2009review,
  title={A review on the Mullins effect},
  author={Diani, Julie and Fayolle, Bruno and Gilormini, Pierre},
  journal={European Polymer Journal},
  volume={45},
  number={3},
  pages={601--612},
  year={2009},
  publisher={Elsevier}
}

\end{document}